\documentclass{aastex701}
\begin{document}

\title{Photometric Redshift PDFs via Neural Network Classification for DESI Legacy Imaging Surveys and Pan-STARRS}

\correspondingauthor{Jun-Qing Xia, Zhong-Lue Wen}
\email{xiajq@bnu.edu.cn, zhonglue@nao.cas.cn}

\author{Da-Chuan Tian}
\affiliation{Institute for Frontiers in Astronomy and Astrophysics, Beijing Normal University, Beijing 100875, People's Republic of China}
\affiliation{School of Physics and Astronomy, Beijing Normal University, Beijing 100875, People's Republic of China}
\email{tiandc@mail.bnu.edu.cn}

\author{Zhong-Lue Wen}
\affiliation{National Astronomical Observatories, Chinese Academy of Sciences, 20A Datun Road, Chaoyang District, Beijing 100101, People's Republic of China}
\email{zhonglue@nao.cas.cn}

\author{Jun-Qing Xia}
\affiliation{Institute for Frontiers in Astronomy and Astrophysics, Beijing Normal University, Beijing 100875, People's Republic of China}
\affiliation{School of Physics and Astronomy, Beijing Normal University, Beijing 100875, People's Republic of China}
\email{xiajq@bnu.edu.cn}

\begin{abstract}
We present a neural network classification (NNC) method for photometric redshift estimation that produces well-calibrated redshift probability density functions (PDFs).
The method discretizes the redshift space into ordered bins and optimizes the Continuous Ranked Probability Score (CRPS), which respects the ordinal nature of redshift and naturally provides uncertainty quantification.
Unlike traditional regression approaches that output single point estimates, our method can capture multi-modal posterior distributions arising from color-redshift degeneracies.
We apply this method to the DESI Legacy Imaging Surveys Data Release 10 (LSDR10) and Pan-STARRS Data Release 2 (PS1DR2), using an unprecedented spectroscopic training sample from DESI DR1 and SDSS DR19.
Our method achieves $\sigma_{\mathrm{NMAD}} = 0.0153$ and $\eta = 0.50\%$ on LSDR10, and $\sigma_{\mathrm{NMAD}} = 0.0222$ and $\eta = 0.34\%$ on PS1DR2 combined with unWISE infrared photometry.
The NNC method outperforms Random Forest, XGBoost, and standard neural network regression.
We demonstrate that DESI DR1 significantly improves photo-$z$ performance at $z > 1$, while the combination of deep optical photometry and mid-infrared coverage is essential for achieving high precision across the full redshift range.
We provide a unified photometric redshift catalog combining LSDR10 and PS1DR2 with a hierarchical model selection strategy based on available photometry.
The well-calibrated PDFs produced by our method are valuable for cosmological studies and can be extended to next-generation surveys such as CSST, Euclid, and LSST.

\end{abstract}


\section{Introduction}

Photometric redshifts (photo-$z$s) are essential for modern extragalactic astronomy and observational cosmology.
While spectroscopic observations provide precise redshift measurements, the observational cost scales prohibitively with sample size, limiting spectroscopic surveys to millions of galaxies.
In contrast, imaging surveys can efficiently observe billions of sources, but determining their distances requires estimating redshifts from broadband photometry alone.
Accurate photo-$z$ estimates are crucial for a wide range of astrophysical and cosmological applications, including galaxy evolution studies \citep{ilbert2013, muzzin2013}, large-scale structure analyses \citep{ross2011, ho2012}, weak gravitational lensing \citep{hildebrandt2017, asgari2021, descollaboration2022}, galaxy cluster detection \citep{wen2012, rykoff2014a, wen2024a, tian2025}, and cosmological parameter constraints \citep{abbott2018, heymans2021}.

Photo-$z$ estimation methods generally fall into two categories: template-fitting and machine learning approaches.
Template-fitting methods \citep[e.g.,][]{arnouts1999, bolzonella2000, benitez2000, ilbert2006, brammer2008} compare observed photometry against libraries of spectral energy distributions to infer redshifts, offering physical interpretability but suffering from template incompleteness and systematic biases when the true galaxy population deviates from the assumed templates.
Machine learning methods \citep[e.g.,][]{collister2004, carliles2010, carrascokind2013, almosallam2016, sadeh2016a} learn empirical mappings between photometric features and spectroscopic redshifts, achieving higher precision when sufficient training data are available, but their performance is fundamentally limited by the size, quality, and representativeness of the spectroscopic training sample.

Traditional machine learning approaches for photo-$z$ estimation---including artificial neural networks \citep[e.g.,][]{firth2003, collister2004}, random forests \citep[e.g.,][]{carrascokind2013, zhou2021}, and gradient boosting methods \citep[e.g.,][]{gerdes2010, li2024}---typically formulate the problem as a regression task, outputting single point estimates.
However, single-value regression cannot provide uncertainty information, capture multi-modal posteriors arising from color-redshift degeneracies, or avoid systematic biases in sparsely sampled redshift regions.
These limitations motivate the development of methods that produce full redshift probability density functions (PDFs), enabling proper uncertainty propagation in downstream analyses.

Recent years have witnessed significant progress in probabilistic photo-$z$ estimation through various approaches, including mixture density networks \citep{disanto2018}, Bayesian neural networks \citep{zhou2022, zhou2024, jones2024}, and classification-based methods that discretize the redshift space into bins \citep{pasquet2019, schuldt2021, lee2021}.
Among these, the binning approach offers a natural framework for PDF estimation while maintaining computational efficiency and interpretability.
Comprehensive comparisons of these PDF-producing methods have been conducted in the context of LSST \citep{schmidt2020, zhang2025, team2026}.

The advent of the Dark Energy Spectroscopic Instrument (DESI) \citep{collaboration2016a} has revolutionized the landscape of spectroscopic training data.
DESI DR1 \citep{collaboration2025h}, released in March 2025, provides highly reliable redshifts for over 13 million galaxies across 9,000 deg$^2$, dramatically expanding the available spectroscopic training sample.
More importantly, DESI extends spectroscopic coverage to intermediate and high redshifts ($z > 0.8$) through its Luminous Red Galaxy (LRG) and Emission Line Galaxy (ELG) target selections, filling critical gaps that previously limited photo-$z$ performance at $z \gtrsim 0.8$.
We note that these target selections probe specific galaxy populations---LRGs are massive, red, passively evolving galaxies, while ELGs are blue, star-forming systems selected via [O{\sc ii}] emission---rather than providing a magnitude-limited census of the full galaxy population at these redshifts (see Section~\ref{sec:discuss_desi} for further discussion of the implications).
This unprecedented dataset provides an ideal opportunity to develop and evaluate improved photo-$z$ methods.

In this work, we present a neural network classification (NNC) method for photometric redshift estimation that leverages the unprecedented spectroscopic sample from DESI DR1 combined with the Sloan Digital Sky Survey Data Release 19 (SDSS DR19).
Our method discretizes the redshift space into ordered bins and trains the network to output probability distributions over these bins.
By optimizing the Continuous Ranked Probability Score (CRPS), the model produces well-calibrated redshift probability density functions.

We apply our method to two major photometric surveys: the DESI Legacy Imaging Surveys Data Release 10 (LSDR10), which provides deep optical photometry in the $g$, $r$, $i$, and $z_{\mathrm{m}}$ bands combined with WISE $W1$ and $W2$ infrared bands, and Pan-STARRS Data Release 2 (PS1DR2), which offers optical photometry in the $g$, $r$, $i$, $z_{\mathrm{m}}$, and $y$ bands over a larger sky area.
This enables a systematic comparison of photo-$z$ performance across surveys with different photometric depths and wavelength coverage.
We further investigate the impact of spectroscopic training sample composition by comparing models trained on SDSS-only, DESI-only, and combined samples.

This paper is organized as follows.
Section~\ref{sec:data} describes the spectroscopic and photometric datasets, followed by the NNC methodology in Section~\ref{sec:methods}.
Section~\ref{sec:results} reports the photo-$z$ performance on LSDR10 and PS1DR2, including comparisons with other methods.
In Section~\ref{sec:discussion}, we discuss the advantages of our approach, the characteristics of DESI DR1, and the performance differences between surveys.
Section~\ref{sec:catalog} describes the construction of a unified photometric redshift catalog combining both surveys.
Finally, Section~\ref{sec:summary} summarizes our main conclusions.

\section{Data} \label{sec:data}
\subsection{Spectroscopic Training Sample}
We collect spectroscopic galaxies from SDSS DR19 \citep{kollmeier2025} and DESI DR1 \citep{collaboration2016a, collaboration2025h}, whose spectroscopic redshifts are regarded as the ground-truth to train the deep learning model for estimating photometric redshifts.

SDSS DR19 is the second data release for the fifth phase of the Sloan Digital Sky Survey (SDSS-V) \citep{kollmeier2025}, comprising more spectroscopic redshifts of galaxies than ever.
The SDSS spectroscopic galaxy samples span several programs with different targeting strategies: the Main Galaxy Sample is a magnitude-limited selection ($r_{\rm Petro} < 17.77$) at $z \lesssim 0.3$ \citep{strauss2002}; BOSS targeted massive galaxies through the LOWZ ($z \sim 0.15$--$0.4$) and CMASS ($z \sim 0.4$--$0.7$) samples using color-magnitude cuts \citep{padmanabhan2012,dawson2012}; and eBOSS extended coverage to LRGs at $z \sim 0.6$--$1.0$ and ELGs at $z \sim 0.7$--$1.1$ \citep{dawson2016,prakash2016,raichoor2017}.
We use SDSS CasJobs\footnote{\url{https://skyserver.sdss.org/CasJobs/}} to query spectroscopic galaxies with $\texttt{zwarningnoqso} = 0$, $\texttt{classnoqso} =
\texttt{`GALAXY'}$, $\texttt{znoqso} > 0$, $\texttt{zErrnoqso} > 0$ for BOSS/eBOSS, and $\texttt{zwarning} = 0$, $\texttt{class} = \texttt{`GALAXY'}$, $\texttt{z} > 0$, $\texttt{zErr} > 0$ for the rest sources.
For BOSS/eBOSS galaxy spectra, the \texttt{*noqso} quantities are the SDSS redshift/classification outputs obtained with quasar templates excluded, which avoids spurious QSO fits when selecting galaxies. Overall, the SDSS cuts retain only confidently classified galaxies with successful redshift measurements and valid redshift uncertainties.

DESI DR1 was officially released in March 2025, containing all the data from the first 13 months of the DESI 4-meter telescope at Kitt Peak National Observatory in the United States, as well as the unified reprocessing of the DESI Survey Validation data previously made public in the DESI Early Data Releases.
The DR1 contains highly reliable redshift data for 18.7 million objects over more than 9,000 deg$^2$, among which 13.1 million have been spectroscopically classified as galaxies, significantly expanding the sample of galaxies with spectroscopic measurements.
The DESI spectroscopic targets relevant to this work include: the BGS, a low-redshift galaxy survey comprising a magnitude-limited bright sample ($r < 19.5$) and a fainter $19.5 < r < 20.175$ component, primarily at $z < 0.6$ \citep{hahn2023}; LRGs, color-selected luminous red galaxies at $0.4 \lesssim z \lesssim 1.0$ using $g$, $r$, $z_{\mathrm{m}}$, and WISE $W1$ photometry \citep{zhou2023a}; and ELGs, color-selected star-forming galaxies over $0.6 \lesssim z \lesssim 1.6$ using $g$, $r$, and $z_{\mathrm{m}}$ photometry, with the selection optimized toward $1.1 \lesssim z \lesssim 1.6$, targeting galaxies with strong [O{\sc ii}] emission \citep{raichoor2023}.
We select all BGS galaxies, LRGs, and ELGs and filter them by $\texttt{SPECTYPE} = \texttt{`GALAXY'}$, $\texttt{ZWARN} = 0$, $\texttt{MASKBITS} = 0$,
$\texttt{MORPHTYPE} \neq \texttt{`PSF'}$, $\texttt{FLUX} > 0$ and $\texttt{FLUXIVAR} > 0$ in $g$, $r$, $z_{\mathrm{m}}$, $\mathrm{W1}$, and $\mathrm{W2}$ bands.
These DESI cuts retain only galaxy spectra with reliable redshift solutions, no imaging-mask contamination, extended morphology, and valid positive flux measurements in the required bands.
In addition, a threshold of $\texttt{DELTACHI2} > 25$ is applied, which quantifies the statistical significance of the best-fitting redshift solution relative to the next best alternative \citep{collaboration2025g}, and serves as an indicator of spectroscopic redshift reliability.
We tested \texttt{DELTACHI2} thresholds ranging from 25 to 5000 and adopted 25 as the optimal value, balancing sample size with spectroscopic quality.

We compile our spectroscopic sample from SDSS DR19 and DESI DR1, requiring $\texttt{zErr} < 0.001$. Using TOPCAT \citep{taylor2005}, we cross-match the two surveys within a $1^{\prime\prime}$ radius. For duplicated sources, we firstly keep the entry with the smaller redshift error, and then exclude those matches where the redshift difference exceeds 0.005. Consequently, we obtain a final spectroscopic catalog containing 11.4 million sources.

\subsection{Photometric Data}
\subsubsection{DESI Legacy Imaging Surveys DR10}
The DESI Legacy Imaging Surveys \citep{dey2019} are a combination of three public imaging projects designed to provide target selection for DESI: the DECam Legacy Survey (DECaLS) covering the southern sky in $g$, $r$, and $z_{\mathrm{m}}$ bands, the Beijing-Arizona Sky Survey (BASS) \citep{zou2017} providing $g$- and $r$-band imaging for the northern sky, and the Mayall $z_{\mathrm{m}}$-band Legacy Survey (MzLS) complementing BASS with $z_{\mathrm{m}}$-band observations.
In addition, mid-infrared photometry from the Wide-field Infrared Survey Explorer (WISE) \citep{wright2010} W1 (3.4 $\rm \mu m$) and W2 (4.6 $\rm \mu m$) bands is incorporated through forced photometry on unWISE coadded images.

LSDR10, released in 2022, is the final data release of the Legacy Surveys, covering approximately 20,000 square degrees including the full DESI footprint. The $5\sigma$ point-source depths reach approximately 24.7, 23.9, and 23.0 mag (AB) in $g$, $r$, and $z_{\mathrm{m}}$ bands, respectively, and approximately 20.4 and 19.5 mag in WISE $W1$ and $W2$ bands.

We retrieved the LSDR10 photometric data from the \texttt{ls\_dr10.tractor} table via the Astro Data Lab platform.
Additionally, we obtained the official photometric redshift estimates from the same source to facilitate subsequent comparative analysis.
The following cleaning criteria are applied to this dataset:
\begin{enumerate}
\item \texttt{type} $\neq$ \texttt{PSF};
\item \texttt{brick\_primary} $=$ 1: selects the primary detection for each source to avoid duplicate entries from overlapping imaging bricks;
\item \texttt{fracflux} $<$ 0.5, \texttt{fracmasked} $<$ 0.4, and \texttt{fracin} $>$ 0.3 for the $g$, $r$, and $z_{\mathrm{m}}$ bands. These quantities characterize contamination from neighboring sources, the fraction of masked pixels, and the fraction of source flux contained within the fitted blob, respectively;
\item Signal-to-noise ratio (\texttt{snr}) $\geq$ 5 for the $g$, $r$, $z_{\mathrm{m}}$, $W1$, and $W2$ bands;
\item Valid magnitude in $g$, $r$, $z_{\mathrm{m}}$, $W1$, and $W2$ bands;
\item Redshift range restricted to $0.001 < z < 2$.
\end{enumerate}

Native LSDR10 $i$-band photometry is available only for a subset of sources, primarily within the DECaLS footprint (dec~$\lesssim 32^{\circ}$). We therefore use it only for the primary six-band model based on $g,r,i,z_{\mathrm{m}},W1,W2$. For sources without usable native LSDR10 $i$-band measurements, we use supplementary models based on the available photometry, as incorporated into the catalog construction in Section~\ref{sec:catalog}.

\subsubsection{Pan-STARRS DR2}
The Panoramic Survey Telescope and Rapid Response System (Pan-STARRS) \citep{chambers2019} is a wide-field astronomical imaging survey conducted by the Institute for Astronomy at the University of Hawaii.
Pan-STARRS1 (PS1) survey, using a 1.8 meter telescope, is a part of Pan-STARRS and observed the entire sky north of declination $-30^{\circ}$ (approximately 30,000 square degrees, or $3\pi$ steradians) in five broadband filters: $g$, $r$, $i$, $z_{\mathrm{m}}$, and $y$, spanning wavelengths from approximately 400 nm to 1000 nm.
PS1DR2 \citep{flewelling2020}, released in January 2019, provides stacked photometry reaching $5\sigma$ point-source depths of approximately 23.3, 23.2, 23.1, 22.3 and 21.3 mag (AB) in the $g$, $r$, $i$, $z_{\mathrm{m}}$, and $y$ bands, respectively.

We retrieve the photometric data of the PS1DR2 sources from the \texttt{ObjectThin} and \texttt{StackObjectThin} tables via the PS1 CasJobs SQL Server, requiring $\texttt{nDetection} \ge 1$ and $\texttt{primaryDetection} = 1$. The latter retains only the primary detection to avoid duplicates from overlapping sky cells.
The following quality cuts are applied to the spectroscopic galaxy sample:
\begin{enumerate}
    \item Valid magnitude measurements across all five bands;
    \item $\texttt{KronMagErr} < 0.2, 0.1, 0.05, 0.1, 0.2$ and $\texttt{ApMagErr} < 0.02, 0.01, 0.006, 0.006, 0.006$ for the $g, r, i, z, y$ bands, respectively. These cuts exclude sources with potentially large photometric uncertainties;
    \item $\texttt{PSFMag} - \texttt{KronMag} \ge 0.1$. This cut excludes point-like sources, since galaxies are more extended than the PSF and therefore have larger PSF--Kron differences;
    \item Redshift range restricted to $0.001 < z < 1.2$.
\end{enumerate}

Galactic dust extinction correction is applied to the PS1DR2 galaxies using the \texttt{dustmaps} Python package \citep{green2018a}.
The SFD E(B-V) values are calculated \citep{schlafly2011} and subsequently applied to the PSF, aperture, and Kron magnitudes.

\subsection{Cross-matched Samples} \label{sec:cross_match}

We cross-match the LSDR10 and PS1DR2 photometric catalogs with the spectroscopic catalog described previously, using TOPCAT with a $1^{\prime\prime}$ matching radius.
Cross-matching the full spectroscopic sample with the primary six-band LSDR10 subset and with PS1DR2 yields 3,963,602 and 6,756,659 matched sources, respectively.
The spectroscopic galaxy sample has been expanded by about three times in the era of DESI.

We additionally construct a PS1DR2 + unWISE sample by augmenting the PS1DR2 photometry with mid-infrared W1 and W2 bands from the unWISE catalog \citep{schlafly2019}.
We cross-match the PS1DR2 sample with unWISE using TOPCAT within a $1^{\prime\prime}$ matching radius, requiring \texttt{flags\_unwise} $= 0$ and \texttt{flags\_info} $= 0$ in both W1 and W2 bands to exclude sources affected by bright star artifacts or other contamination, and further requiring valid magnitude measurements with positive flux errors.
After applying these criteria, we obtain $\sim$4.2 million training sources, $\sim$520,000 validation sources, and $\sim$520,000 test sources.
The input features for this configuration are expanded from 30 to 34, adding W1 and W2 magnitudes and their errors.

Stellar masses used in the analysis are estimated following the methods of \citet{wen2020} and \citet{wen2024a}.

\begin{figure}[ht!]
\centering
\includegraphics[width=\textwidth]{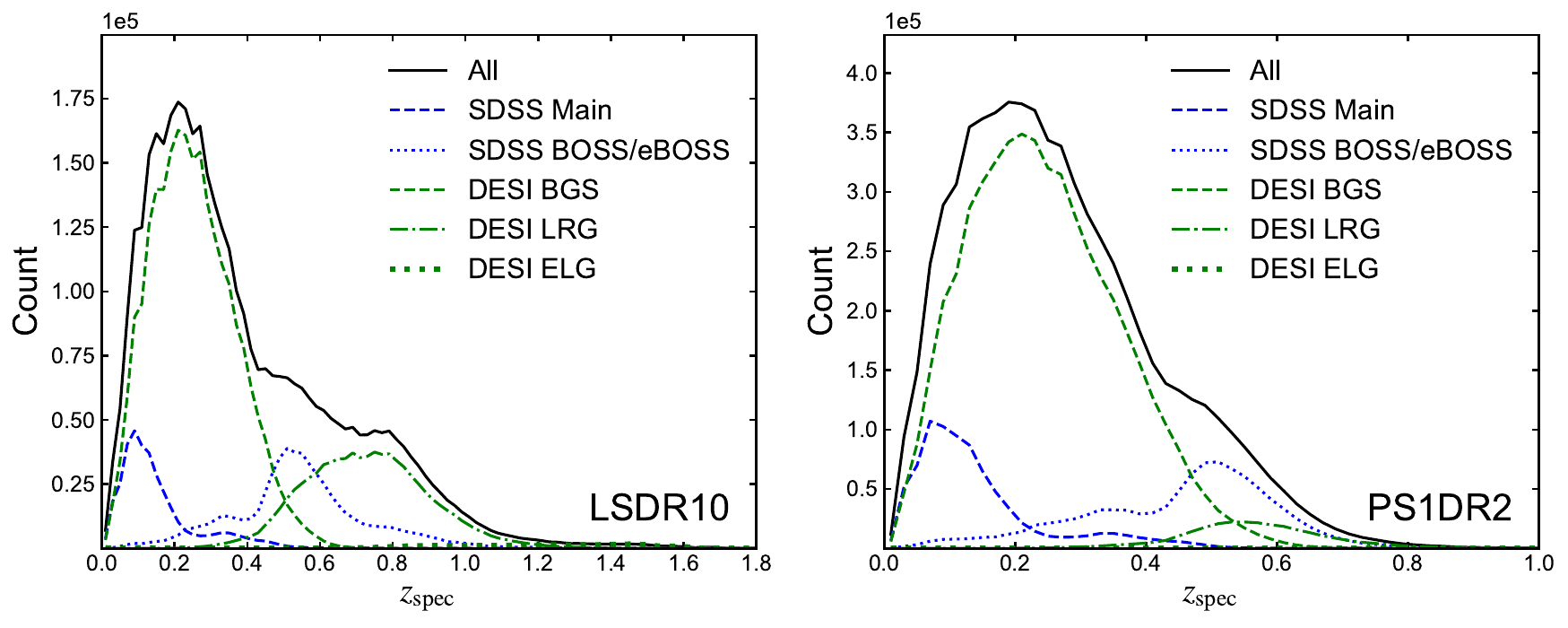}
\caption{Redshift distributions of the spectroscopic training samples for LSDR10 (left) and PS1DR2 (right), broken down by spectroscopic source. In each panel, the black line shows all sources, while the colored lines indicate the SDSS Main (blue dashed), SDSS BOSS/eBOSS (blue dotted), DESI BGS (green solid), DESI LRG (green dash-dotted), and DESI ELG (green dotted) subsamples. The DESI ELG distribution may be less visually prominent because the matched ELG subsample is much smaller than the BGS and LRG subsamples.}
\label{fig:sample_distribution}
\end{figure}

The combined training samples for LSDR10 and PS1DR2 are then divided into training, validation, and test sets with a split ratio of 8:1:1.
The training set is used to optimize model parameters, while the validation set is employed for hyperparameter tuning.
The test set, reserved as a completely independent and unseen sample, is used for the final evaluation of model performance, enabling a fair comparison across different methods and model configurations.
Figure~\ref{fig:sample_distribution} shows the redshift distributions of the spectroscopic training samples for LSDR10 and PS1DR2. 

\subsection{Spectroscopic Sample Properties} \label{sec:spectro_properties}

\begin{figure*}[ht!]
\centering
\includegraphics[width=\textwidth]{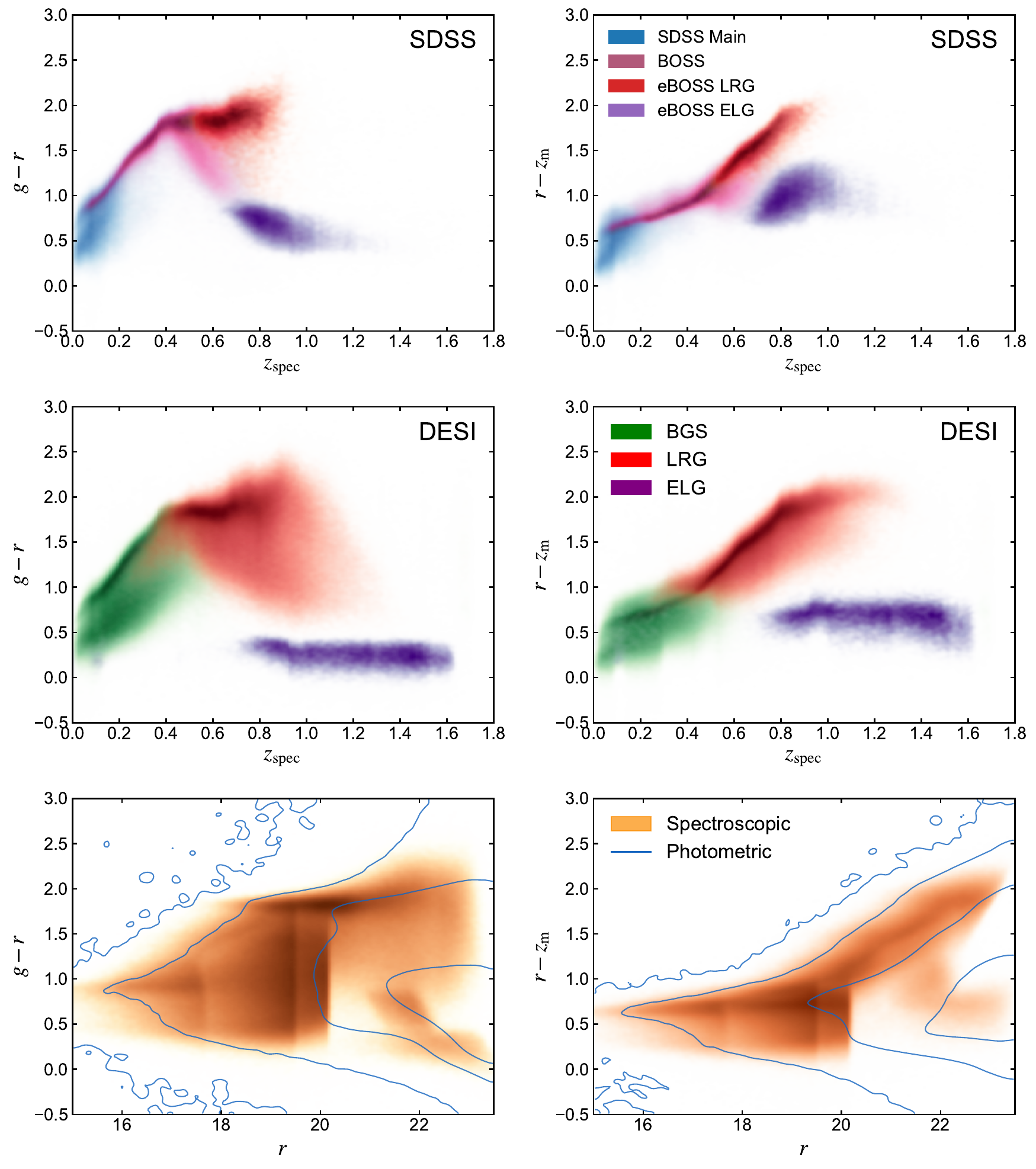}
\caption{Comparison of spectroscopic-sample coverage in LSDR10. Top row: color--redshift distributions for the SDSS spectroscopic sample, separated into the SDSS Main (blue), BOSS (pink), eBOSS LRG (red), and eBOSS ELG (purple) components. Middle row: color--redshift distributions for the DESI spectroscopic sample, color-coded by target type: BGS (green), LRG (red), and ELG (purple). Left and right columns show $g-r$ and $r-z_{\mathrm{m}}$ versus spectroscopic redshift, respectively. Bottom row: color--magnitude distributions in LSDR10, comparing the spectroscopic sample (orange density map) with a random reference sample of $10^7$ photometric sources (blue contours). These panels illustrate how well the spectroscopic sample covers the main photometric locus in color--magnitude space.}
\label{fig:color_redshift_comparison}
\end{figure*}

Figure~\ref{fig:color_redshift_comparison} summarizes the redshift, color, and magnitude coverage of the spectroscopic training samples in LSDR10. The top row separates the SDSS spectroscopic sample into four broad components: the SDSS Main sample, the BOSS galaxy sample, and the eBOSS LRG and ELG subsamples. The SDSS sample is predominantly confined to $z < 0.8$, with the Main Galaxy Sample concentrated at low redshift, the BOSS galaxies populating the intermediate-redshift regime, and the eBOSS LRG and ELG subsamples extending the SDSS coverage to higher redshift with distinct loci in the $g-r$ and $r-z_{\mathrm{m}}$ color spaces. The middle row presents the corresponding DESI BGS, LRG, and ELG populations in the same color--redshift planes, enabling a direct visual comparison between the SDSS and DESI spectroscopic selections. At low redshifts ($0.1 < z < 0.4$), BGS reaches $\sim$1.7 mag fainter than the SDSS main galaxy sample \citep{strauss2002, hahn2023}, providing denser coverage for intermediate-luminosity galaxies. The bottom panels compare the color--magnitude distributions of the spectroscopic sample with a random reference sample of $10^7$ photometric LSDR10 sources. These panels show that the spectroscopic sample broadly traces the main photometric locus, but its coverage becomes increasingly incomplete toward the faint and sparsely populated regions of color--magnitude space. Together, these trends motivate our investigation of training sample composition in Section~\ref{sec:training_sample}, while the implications of the targeted nature of these samples for photo-$z$ performance are discussed in Section~\ref{sec:discuss_desi}.

\section{Methods} \label{sec:methods}
\subsection{Neural Network Classification}
In this work, we adopt an NNC approach for photometric redshift estimation.
Unlike traditional neural networks that directly model photo-z as a regression problem with a single point estimate output, our approach discretizes the continuous redshift space into ordered bins and trains the network to classify galaxies into these bins, outputting a probability distribution that enables the estimation of the full redshift PDF.

Traditional ANN regression methods suffer from several limitations:
(1) they provide no uncertainty information for the redshift estimate;
(2) they cannot capture the multi-modal nature of the redshift posterior distribution arising from color-redshift degeneracies;
and (3) they tend to exhibit systematic bias toward lower redshifts in sparsely sampled high-redshift regions, as the loss function is dominated by the densely populated low-redshift samples.

The NNC method addresses these issues by partitioning the redshift range $[z_{\min}, z_{\max}]$ into $N_{\mathrm{bin}}$ equally-spaced bins.
The network outputs an $N_{\mathrm{bin}}$-dimensional probability vector $\mathbf{p} = (p_1, p_2, \ldots, p_{N_{\mathrm{bin}}})$, where $p_i$ represents the probability that the target redshift falls within the $i$-th bin, satisfying $\sum_i p_i = 1$.

Compared to direct regression, the binning approach offers several advantages:
(1) it constrains the output to the valid redshift range $[z_{\min}, z_{\max}]$, avoiding negative or unrealistic redshift predictions;
(2) it provides complete redshift PDFs with built-in uncertainty quantification;
(3) it naturally accommodates multi-modal posteriors arising from color-redshift degeneracies;
and (4) combined with proper scoring rules, the binning formulation produces well-calibrated probability estimates.

We employ a multi-layer perceptron with residual connections as the backbone network.
The architecture consists of four fully-connected hidden layers with dimensions [512, 256, 128, 64], each followed by batch normalization, ReLU activation, and dropout with a rate of 0.1.
Residual connections \citep{he2016} are incorporated between layers to enhance information flow and prevent the degradation of feature representations.
The final output layer produces logits, which are transformed into a probability distribution via the softmax function.

The model is trained by minimizing the CRPS \citep{hersbach2000}, a proper scoring rule that evaluates the quality of probabilistic forecasts. The CRPS is defined as:
\begin{equation}
\mathrm{CRPS}(F, z_{\mathrm{spec}}) = \int_{-\infty}^{\infty} \left[ F(z) - \mathbf{1}(z \geq z_{\mathrm{spec}}) \right]^2 \mathrm{d}z
\end{equation}
where $F(z) = \sum_{i: z_i^{\mathrm{center}} \leq z} p_i$ is the predicted cumulative distribution function, and $\mathbf{1}(z \geq z_{\mathrm{spec}})$ is the Heaviside step function at the true spectroscopic redshift. For the discretized redshift bins, the CRPS can be efficiently computed as:
\begin{equation}
\mathrm{CRPS} = \sum_{i=1}^{N_{\mathrm{bin}}} \left[ F_i - \mathbf{1}(z_i^{\mathrm{center}} \geq z_{\mathrm{spec}}) \right]^2 \Delta z
\end{equation}
where $F_i = \sum_{j=1}^{i} p_j$ is the cumulative probability up to bin $i$, and $\Delta z$ is the bin width.
The CRPS loss is particularly well-suited for photometric redshift estimation for two key reasons.
First, unlike cross-entropy which treats redshift bins as independent labels, CRPS explicitly accounts for the ordinal nature of redshift by operating on the CDF.
This allows the model to penalize errors in proportion to their distance from the true value, effectively distinguishing between minor deviations and catastrophic failures.
Second, as a strictly proper scoring rule, CRPS inherently balances sharpness and calibration.
It drives the model to produce probability distributions that are as narrow (sharp) as possible while maintaining statistical reliability, effectively preventing both under-confident (overly broad) and over-confident (misplaced) predictions.

Driven by CRPS optimization, the model learns to produce probability distributions that are tightly constrained around the predicted redshift.
This morphological behavior directly reflects the model's confidence: for galaxies with distinct photometric signatures, the PDF exhibits a single, sharp peak (unimodal) with minimal probability mass scattered elsewhere.
Conversely, for sources subject to color-redshift degeneracies, the PDF transitions to a broader or multi-modal distribution, naturally encoding the inherent uncertainty in the estimate.

During inference, the redshift point estimate is obtained as the probability-weighted expectation:
\begin{equation}
z_{\mathrm{phot}} = \sum_{i=1}^{N_{\mathrm{bin}}} p_i \cdot z_i^{\mathrm{center}}
\end{equation}
where $z_i^{\mathrm{center}}$ is the central redshift of the $i$-th bin. The full probability distribution also supports Monte Carlo sampling for downstream analyses requiring photo-z PDFs.

We partition the redshift range $[0, 2]$ into $N_{\mathrm{bin}} = 400$ equally-spaced bins for LSDR10 and PS1DR2, which exceeds the redshift coverage of two surveys to ensure complete coverage of the target distribution.

\subsection{Feature Selection and Training Strategy}
For a fair comparison between LSDR10 and PS1DR2, we use magnitudes and magnitude errors as the input features for both surveys.
Specifically, for LSDR10, we use the $g$, $r$, $i$, $z_{\mathrm{m}}$, $W1$, and $W2$ band magnitudes along with their corresponding errors (12 features in total).
For PS1DR2, we use the $g$, $r$, $i$, $z_{\mathrm{m}}$, and $y$ band magnitudes measured in three aperture types (Kron, PSF, and Aperture) together with their errors (30 features in total). 
Unlike the Tractor model-fitting photometry used by LSDR10 \citep{dey2019}, PS1DR2 provides independent measurements through different aperture definitions. Including all three types allows the model to exploit the implicit morphological information encoded in their differences, such as source size and concentration.
We systematically evaluated different feature combinations and found that adding color features provides negligible improvement, so we adopt the magnitude-only configuration for consistency.

We use the AdamW optimizer \citep{loshchilov2019} with an initial learning rate of $5 \times 10^{-4}$ and weight decay of $1 \times 10^{-5}$. The learning rate is reduced by a factor of 0.5 when the validation loss plateaus for 5 consecutive epochs.
Gradient clipping with a maximum norm of 1.0 is applied to stabilize training.
Early stopping with a patience of 20 epochs is employed to prevent overfitting.
The model is trained with a batch size of 4096 for up to 300 epochs.

\subsection{Performance Metrics}
We evaluate the photometric redshift performance using several standard metrics. The normalized residual is defined as:
\begin{equation}
\Delta z_{\mathrm{norm}} = \frac{z_{\mathrm{phot}} - z_{\mathrm{spec}}}{1 + z_{\mathrm{spec}}}
\end{equation}
The outlier fraction ($\eta$) measures the percentage of catastrophic failures, defined as the fraction of sources with $|\Delta z_{\mathrm{norm}}| > 0.15$:
\begin{equation}
\eta = \frac{N(|\Delta z_{\mathrm{norm}}| > 0.15)}{N_{\mathrm{total}}} \times 100
\end{equation}
To prevent these catastrophic outliers from dominating the statistics, bias and $\sigma$ are computed using only non-outlier sources ($|\Delta z_{\mathrm{norm}}| \leq 0.15$).
The bias quantifies the systematic offset, defined as the mean of the normalized residuals for non-outlier sources:
\begin{equation}
\mathrm{bias} = \langle \Delta z_{\mathrm{norm}} \rangle_{\mathrm{non\text{-}outlier}}
\end{equation}
The standard deviation ($\sigma$) measures the dispersion of the normalized residuals, also computed for non-outlier sources only:
\begin{equation}
\sigma = \mathrm{std}(\Delta z_{\mathrm{norm}})_{\mathrm{non\text{-}outlier}}
\end{equation}
The normalized median absolute deviation ($\sigma_{\mathrm{NMAD}}$) provides a robust measure of scatter, computed using all sources:
\begin{equation}
\sigma_{\mathrm{NMAD}} = 1.4826 \times \mathrm{median} \left( \left| \Delta z_{\mathrm{norm}} - \mathrm{median}(\Delta z_{\mathrm{norm}}) \right| \right)
\end{equation}

These metrics together characterize the reliability ($\eta$), accuracy (bias), and precision ($\sigma$, $\sigma_{\mathrm{NMAD}}$) of photometric redshift estimates.

\section{Results} \label{sec:results}

In this section, we present the photometric redshift results from applying our NNC method to LSDR10 and PS1DR2.
Section~\ref{sec:overall_performance} reports the overall performance across different configurations and training samples.
Section~\ref{sec:lsdr10_pdfs} demonstrates the redshift PDFs and calibration diagnostics.
Section~\ref{sec:comparison_ml} compares our method with other approaches.

\subsection{Overall Performance} \label{sec:overall_performance}

\subsubsection{LSDR10 Performance} \label{sec:lsdr10_performance}

We first apply the NNC method to LSDR10, training on the spectroscopic sample described in Section~\ref{sec:data}.
The model achieves consistent performance across training, validation, and test sets, indicating no overfitting.
Below we report results on the independent test set.

\begin{figure*}[ht!]
\centering
\includegraphics[width=\textwidth]{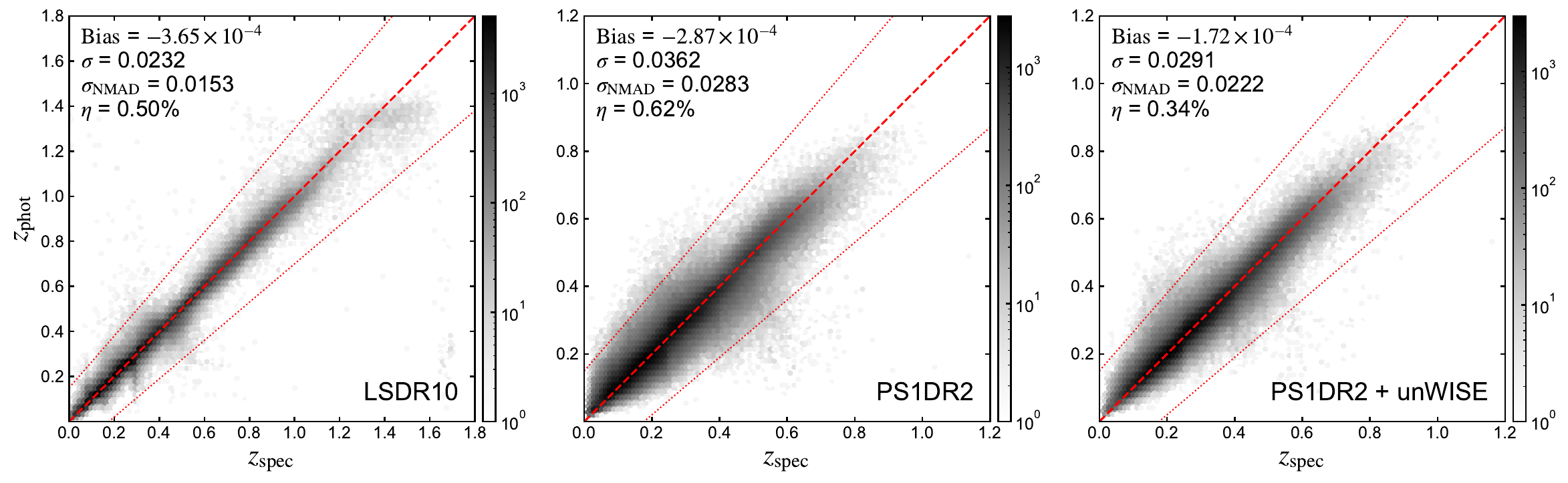}
\caption{Comparison of photometric redshifts ($z_{\mathrm{phot}}$) and spectroscopic redshifts ($z_{\mathrm{spec}}$) for the test sets. Left: LSDR10; Middle: PS1DR2 (optical only); Right: PS1DR2 + unWISE. The red dashed lines indicate the one-to-one relation, and the red dotted lines show the outlier boundaries ($|\Delta z_{\mathrm{norm}}| = 0.15$). Performance metrics are shown in each panel.}
\label{fig:overall_performance}
\end{figure*}

The left panel of Figure~\ref{fig:overall_performance} shows the comparison between photometric and spectroscopic redshifts for LSDR10.
The test set achieves $\sigma_{\mathrm{NMAD}} = 0.0153$, $\eta = 0.50\%$, and $\mathrm{bias} = -3.65 \times 10^{-4}$, demonstrating excellent photo-$z$ precision with low catastrophic failure rates.
The performance metrics are summarized in Table~\ref{tab:table1}.

\begin{deluxetable*}{lcccc}
\tablecaption{Comprehensive Photometric Redshift Performance Summary \label{tab:table1}}
\tablehead{
\colhead{Configuration} & \colhead{Bias} & \colhead{$\sigma$} & \colhead{$\sigma_{\mathrm{NMAD}}$} & \colhead{$\eta$ (\%)}
}
\startdata
\multicolumn{5}{c}{LSDR10} \\
\hline
\textbf{Full (testset)} & \textbf{$-3.65 \times 10^{-4}$} & \textbf{0.0232} & \textbf{0.0153} & \textbf{0.50} \\
$z_{\mathrm{m}} < 21$ & $-4.43 \times 10^{-4}$ & 0.0223 & 0.0151 & 0.36 \\
$z_{\mathrm{m}} \geq 21$ & $3.86 \times 10^{-3}$ & 0.0505 & 0.0449 & 7.29 \\
$\log M_\star \geq 11$ & $6.92 \times 10^{-4}$ & 0.0203 & 0.0136 & 0.25 \\
$10 \leq \log M_\star < 11$ & $-7.29 \times 10^{-4}$ & 0.0233 & 0.0157 & 0.43 \\
$\log M_\star < 10$ & $-1.89 \times 10^{-3}$ & 0.0244 & 0.0174 & 0.55 \\
BGS & $-1.00 \times 10^{-3}$ & 0.0230 & 0.0155 & 0.31 \\
LRG & $3.52 \times 10^{-4}$ & 0.0225 & 0.0171 & 0.42 \\
ELG & $5.92 \times 10^{-3}$ & 0.0597 & 0.0689 & 10.97 \\
$g,r,z_{\mathrm{m}},W1,W2$ + PS1DR2 $i$ & $-4.87 \times 10^{-4}$ & 0.0245 & 0.0164 & 0.34 \\
$g,r,z_{\mathrm{m}},W1,W2$ (no $i$) & $-4.24 \times 10^{-4}$ & 0.0282 & 0.0189 & 0.58 \\
SDSS-only train & $-1.98 \times 10^{-3}$ & 0.0259 & 0.0175 & 0.73 \\
DESI-only train & $-4.82 \times 10^{-4}$ & 0.0235 & 0.0157 & 0.51 \\
\hline
\multicolumn{5}{c}{Other ML Methods on LSDR10} \\
\hline
RF & $-7.60 \times 10^{-4}$ & 0.0355 & 0.0278 & 0.64 \\
XGBoost & $-5.88 \times 10^{-4}$ & 0.0290 & 0.0220 & 0.58 \\
ANN & $-3.65 \times 10^{-4}$ & 0.0243 & 0.0170 & 0.49 \\
\hline
\multicolumn{5}{c}{PS1DR2} \\
\hline
Optical-only & $-2.87 \times 10^{-4}$ & 0.0362 & 0.0283 & 0.62 \\
\textbf{Optical + unWISE} & \textbf{$-1.72 \times 10^{-4}$} & \textbf{0.0291} & \textbf{0.0222} & \textbf{0.34} \\
SDSS-only train & $-1.20 \times 10^{-3}$ & 0.0319 & 0.0245 & 0.48 \\
DESI-only train & $3.44 \times 10^{-4}$ & 0.0294 & 0.0224 & 0.34 \\
\enddata
\end{deluxetable*}

Photo-$z$ performance varies systematically with source properties.
Brighter galaxies ($z_\mathrm{m} < 21$) achieve $\sigma_{\mathrm{NMAD}} = 0.0151$, while fainter sources ($z_\mathrm{m} \geq 21$) show degraded performance with $\sigma_{\mathrm{NMAD}} = 0.0449$ and higher outlier fraction ($\eta = 7.29\%$).
Similarly, massive galaxies ($\log M_{\star} \geq 11$) exhibit excellent precision ($\sigma_{\mathrm{NMAD}} = 0.0136$), while low-mass galaxies ($\log M_{\star} < 10$) show higher scatter ($\sigma_{\mathrm{NMAD}} = 0.0174$) and outlier fraction ($\eta = 0.55\%$).
Among DESI target types, BGS and LRGs achieve $\sigma_{\mathrm{NMAD}} \sim 0.015$--$0.017$ with $\eta < 0.5\%$, while ELGs exhibit substantially larger scatter ($\sigma_{\mathrm{NMAD}} = 0.0689$, $\eta = 10.97\%$).
The poor performance on ELGs likely reflects a combination of factors. In the test set, ELGs are typically fainter than BGS and LRGs, with a median $z_{\mathrm{m}}$ magnitude of 21.779 compared to 18.515 for BGS and 19.871 for LRGs. They also show lower median signal-to-noise ratios in the $r$, $i$, $z_{\mathrm{m}}$, $W1$, and $W2$ bands than BGS and LRGs, in addition to the intrinsic color-redshift degeneracies of star-forming populations.
Further improving the accuracy for ELGs likely requires incorporating image information to break these degeneracies \citep{wei2025}.

Because native LSDR10 $i$-band photometry is not available for all sources, we train two additional models to extend the analysis beyond the primary six-band sample.
For sources without usable native LSDR10 $i$-band measurements but with successful PS1DR2 cross-matches, we supplement the LSDR10 $g,r,z_{\mathrm{m}}$ and WISE $W1,W2$ photometry with PS1DR2 $i$-band magnitudes, achieving $\sigma_{\mathrm{NMAD}} = 0.0164$ and $\eta = 0.34\%$.
For sources without usable native LSDR10 $i$-band measurements or PS1DR2 $i$-band matches, we train a five-band model using only $g,r,z_{\mathrm{m}},W1,W2$, which yields $\sigma_{\mathrm{NMAD}} = 0.0189$ and $\eta = 0.58\%$.
These supplementary models enable comprehensive photo-$z$ coverage across the full LSDR10 footprint and are employed in the catalog construction described in Section~\ref{sec:catalog}.

\subsubsection{PS1DR2 Performance} \label{sec:ps1dr2_performance}

We then apply the NNC method to PS1DR2, which covers a larger sky area ($\sim$30,000 deg$^2$) but with shallower photometric depth compared to LSDR10.
Using the five-band $g,r,i,z_{\mathrm{m}},y$ photometry, the model achieves $\sigma_{\mathrm{NMAD}} = 0.0283$ and $\eta = 0.62\%$, as shown in the middle panel of Figure~\ref{fig:overall_performance}.
The scatter is notably larger than LSDR10, suggesting that additional wavelength coverage may be needed to break color-redshift degeneracies.

To test whether infrared photometry can improve PS1DR2 photo-$z$ performance, we augment the PS1DR2 photometry with unWISE W1 and W2 bands, as described in Section~\ref{sec:cross_match}.

As shown in the right panel of Figure~\ref{fig:overall_performance}, adding infrared bands significantly improves performance to $\sigma_{\mathrm{NMAD}} = 0.0222$ and $\eta = 0.34\%$, representing a $\sim$22\% reduction in scatter and nearly 50\% reduction in outlier fraction.
This improvement shows that mid-infrared coverage provides useful indirect constraints through the long optical-to-infrared color baseline, helping improve photo-$z$ precision, particularly for red galaxies at intermediate redshifts.

Despite the improvement, PS1DR2 + unWISE still yields a $\sigma_{\mathrm{NMAD}}$ approximately 1.45 times larger than LSDR10, suggesting that photometric depth is as critical as wavelength coverage.
This performance gap and the specific impact of survey depth are further analyzed in Section~\ref{sec:discuss_photometry}.

\subsubsection{Impact of Training Sample Composition} \label{sec:training_sample}

The composition of the spectroscopic training sample may significantly affect photo-$z$ performance, particularly at high redshifts where DESI provides unprecedented spectroscopic coverage.
To quantify this effect, we train separate models using SDSS-only, DESI-only, and all (SDSS+DESI) spectroscopic samples.
Note that the All model corresponds to the results reported in Sections~\ref{sec:lsdr10_performance} and \ref{sec:ps1dr2_performance}.

For LSDR10, the SDSS-only trained model achieves $\sigma_{\mathrm{NMAD}} = 0.0175$ and $\eta = 0.73\%$, while the DESI-only model yields $\sigma_{\mathrm{NMAD}} = 0.0157$ and $\eta = 0.51\%$.
The All model achieves the best performance ($\sigma_{\mathrm{NMAD}} = 0.0153$, $\eta = 0.50\%$), benefiting from the complementary coverage of SDSS at low redshifts and DESI at intermediate-to-high redshifts.

For PS1DR2 + unWISE, a similar pattern emerges: the SDSS-only model achieves $\sigma_{\mathrm{NMAD}} = 0.0245$ and $\eta = 0.48\%$, while the DESI-only model yields $\sigma_{\mathrm{NMAD}} = 0.0224$ and $\eta = 0.34\%$, comparable to the All result.
These results demonstrate that DESI DR1 substantially improves photo-$z$ performance, while the combination with SDSS provides additional gains through more complete sampling in color-magnitude space.
This performance improvement at high redshift is directly connected to DESI's targeting strategy: DESI ELGs extend the spectroscopic coverage of star-forming galaxies to $0.6 \lesssim z \lesssim 1.6$, with the selection optimized toward $1.1 \lesssim z \lesssim 1.6$; the highest-redshift part of this population is weakly represented or largely absent from SDSS spectroscopy. DESI LRGs provide complementary dense coverage of red galaxies to $z \sim 1.2$ in our matched sample (Figure~\ref{fig:color_redshift_comparison}). The SDSS-only model's sharp degradation at $z > 1.0$ (Figure~\ref{fig:nmad_vs_z}) is consistent with the reduced spectroscopic coverage of these populations in its training data.

\subsubsection{Redshift-Dependent Performance} \label{sec:redshift_dependent}

\begin{figure}[ht!]
\centering
\includegraphics[width=\textwidth]{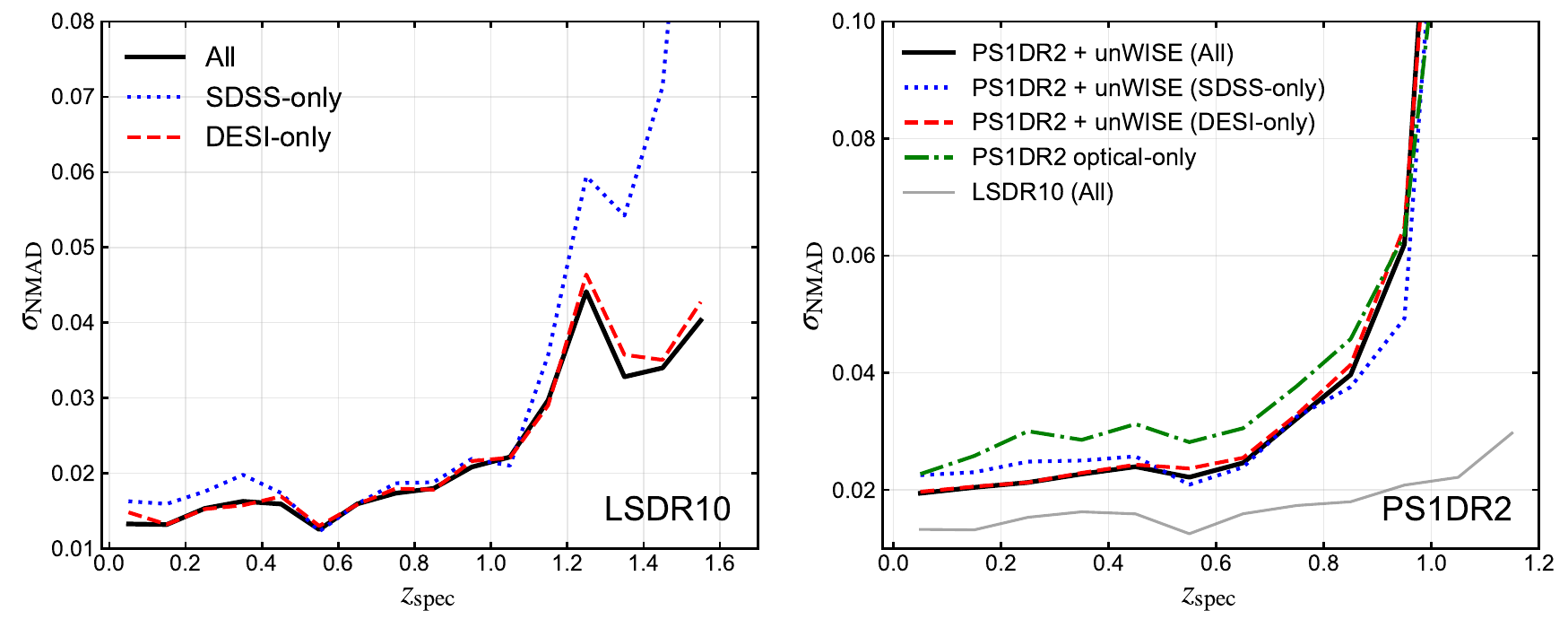}
\caption{$\sigma_{\mathrm{NMAD}}$ as a function of spectroscopic redshift. Left: LSDR10 with different training configurations (All, SDSS-only, DESI-only). Right: PS1DR2 configurations (optical-only; +unWISE with All, SDSS-only, and DESI-only training; and the LSDR10 All model for reference). The comparison demonstrates the impact of training sample composition and infrared photometry.}
\label{fig:nmad_vs_z}
\end{figure}

The overall metrics reported above are averaged over the full redshift range and may obscure performance variations at different redshifts.
To examine redshift-dependent behavior, Figure~\ref{fig:nmad_vs_z} shows $\sigma_{\mathrm{NMAD}}$ as a function of spectroscopic redshift for all survey and training configurations.

For LSDR10 with the All training sample, $\sigma_{\mathrm{NMAD}}$ remains stable at $\lesssim 0.02$ across $0 < z < 0.9$, then gradually increases toward higher redshifts, reaching $\sigma_{\mathrm{NMAD}} \approx 0.04$ at $z \gtrsim 1.2$.
The impact of training sample composition becomes pronounced at high redshifts: the SDSS-only trained model diverges sharply at $z > 1.1$ due to the lack of high-redshift training data, while both the All and DESI-only models maintain $\sigma_{\mathrm{NMAD}} \lesssim 0.05$ out to $z \sim 1.5$.
This clearly demonstrates the critical contribution of DESI DR1 spectroscopy at high redshifts.

Among the PS1DR2 + unWISE models, the SDSS-only trained configuration exhibits systematically larger scatter than the DESI-only and All models, especially at $z \gtrsim 0.8$, while the DESI-only curve closely tracks the All result over most of the redshift range.
For PS1DR2, adding unWISE infrared photometry reduces scatter at $z < 0.85$, but all configurations show steep increases at $z > 0.8$ due to the shallower photometric depth limiting precise color measurements for faint, high-redshift sources.
The comparison between LSDR10 and PS1DR2 across the full redshift range highlights that both deep photometry and broad wavelength coverage are essential for achieving consistently high photo-$z$ precision.

\subsection{Redshift PDFs and Calibration} \label{sec:lsdr10_pdfs}

A key advantage of our NNC method is its ability to produce full redshift PDFs rather than single point estimates.
To demonstrate the detailed probabilistic performance, we present the analysis based on the LSDR10 as a representative example.

\begin{figure*}[ht!]
\centering
\includegraphics[width=\textwidth]{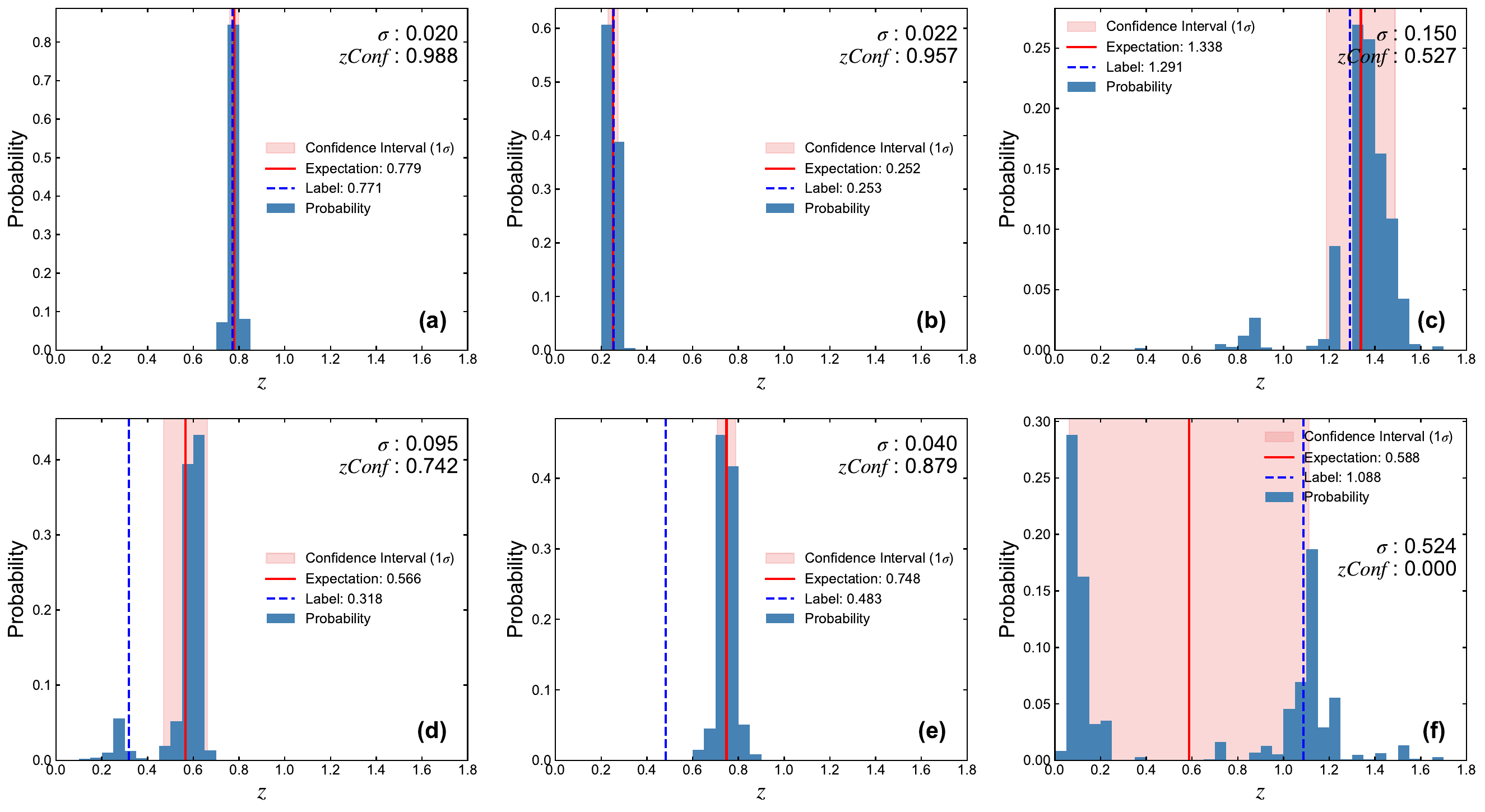}
\caption{Representative examples of redshift PDFs from the LSDR10 test set. Each panel shows the predicted probability distribution as a histogram over 40 redshift bins obtained by rebinning the native 400-bin output. Upper row (a--c): high-confidence predictions where the spectroscopic redshift (blue vertical dashed line) agrees well with the PDF expectation (red vertical solid line). Lower row (d--f): outlier cases showing various failure modes including peaked predictions at incorrect redshifts and broader distributions. The shaded regions indicate the $1\sigma$ confidence intervals.}
\label{fig:pdf_examples}
\end{figure*}

Figure~\ref{fig:pdf_examples} illustrates six representative PDF examples from the test set.
For visualization clarity, each PDF in Figure~\ref{fig:pdf_examples} is shown as a histogram over 40 redshift bins obtained by rebinning the native 400-bin model output. All performance metrics reported in this work are computed from the native 400-bin output. We use the rebinned 40-bin representation only for visualization and public release, because it is smoother and more practical while preserving the overall PDF shape.
The blue dashed line indicates the spectroscopic redshift, the red solid line marks the photometric redshift (expectation value), and the red shaded region represents the $1\sigma$ confidence interval.
For each prediction, we define $zConf$ as the integrated probability within the interval $[z_{\mathrm{exp}} - \alpha(1+z_{\mathrm{exp}}), z_{\mathrm{exp}} + \alpha(1+z_{\mathrm{exp}})]$, where $z_{\mathrm{exp}}$ is the expectation value of the PDF and $\alpha = 0.03$ \citep{carrascokind2013, luo2024}.

The upper row displays three high-quality predictions representing ideal cases.
Panels~(a) and (b) show very sharp PDFs with small uncertainty ($\sigma = 0.020$ and $0.022$) and $zConf$ approaching unity ($0.988$ and $0.957$). In these cases, the red expectation line nearly coincides with the blue true redshift line, indicating both high confidence and high accuracy.
Panel~(c) exhibits a multi-modal distribution peaking at $z \sim 1.3-1.4$. Notably, the spectroscopic redshift lies within the $1\sigma$ interval, confirming the model's ability to robustly quantify uncertainty even when a precise redshift cannot be determined.
These examples demonstrate that the model effectively extracts photometric features and provides reliable probability estimates under typical conditions.

The lower row illustrates challenging scenarios where photometric confusion leads to complex PDF morphologies:
\begin{itemize}
    \item Panel~(d) shows a bi-modal distribution caused by source blending. The optical image reveals two overlapping galaxies: a reddish component and a fainter, bluer component. This contamination results in a composite photometric signal. Consequently, the model assigns the primary probability peak to a higher redshift ($z \sim 0.6$), biasing the prediction toward the redder source, while the true spectroscopic redshift ($z=0.318$) aligns with the secondary, lower-probability peak.
    \item Panel~(e) presents a catastrophic outlier despite high confidence ($zConf=0.879$). The source is a red galaxy in close proximity to another galaxy. Although DESI spectra confirm both galaxies are at similar redshifts ($z \approx 0.48$), their distinct colors likely confounded the model. The model appears to have misidentified the source's spectral type or suffered from photometric contamination from the neighbor, leading to an overestimation of the redshift at $z \sim 0.75$.
    \item Panel~(f) illustrates a complex case driven by color-redshift degeneracy and likely source fragmentation. Although the official catalog reports two photometric sources ($z_{phot} \approx 0.99$ and $0.70$), the existence of a single spectroscopic redshift ($z=1.088$) suggests a single irregular galaxy was incorrectly split by the processing pipeline. This photometric ambiguity results in a bimodal PDF. Crucially, while the point estimate ($z_{exp}=0.588$) fails catastrophically, the full PDF remains robust, correctly capturing the true redshift within the high-$z$ mode despite the morphological confusion.
\end{itemize}

Notably, panels (a), (b), (d), and (e) all exhibit high $zConf$ values yet yield dramatically different outcomes—accurate predictions in (a) and (b) versus significant deviations in (d) and (e).
This discrepancy underscores that $zConf$ primarily reflects the precision of the PDF rather than the actual accuracy of the prediction.
Therefore, high $zConf$ implies the model is internally confident, but it does not guarantee correctness, especially in the presence of anomalies like blending.

In summary, while the PDF shape generally provides a robust measure of uncertainty, external physical factors such as source blending and confusion can occasionally induce overconfidence, leading to catastrophic outliers where the model is ``confidently wrong.''

To assess the statistical calibration of our PDFs, we employ the Probability Integral Transform (PIT) diagnostic \citep{dawid1984, disanto2018, desprez2020, mucesh2021}, following the framework established by \citet{schmidt2020} for systematic comparison of photo-$z$ PDF methods.
For well-calibrated PDFs, the PIT values---defined as the cumulative probability at the true redshift---should follow a uniform distribution \citep{dawid1984}.
The shape of the PIT distribution reveals systematic biases: PIT values concentrated near 0 indicate systematic overestimation of redshifts, while concentration near 1 indicates systematic underestimation.
A U-shaped distribution suggests overconfident predictions (PDFs too narrow), whereas an inverted U-shape indicates conservative predictions (PDFs too broad).
To improve calibration, we apply temperature scaling \citep{guo2017a}, a simple post-hoc calibration method.
The network logits $z$ are divided by a learned temperature parameter $T > 0$ before the softmax operation:
\begin{equation}
p_i = \frac{\exp(z_i / T)}{\sum_j \exp(z_j / T)}
\end{equation}
When $T > 1$, the output distribution becomes softer (broader PDFs), while $T < 1$ produces sharper distributions.
The optimal temperature is determined by minimizing the negative log-likelihood on the validation set.
Unless otherwise noted, all results presented in this work are based on the temperature-scaled calibrated PDFs.

\begin{figure}[ht!]
\centering
\includegraphics[width=\textwidth]{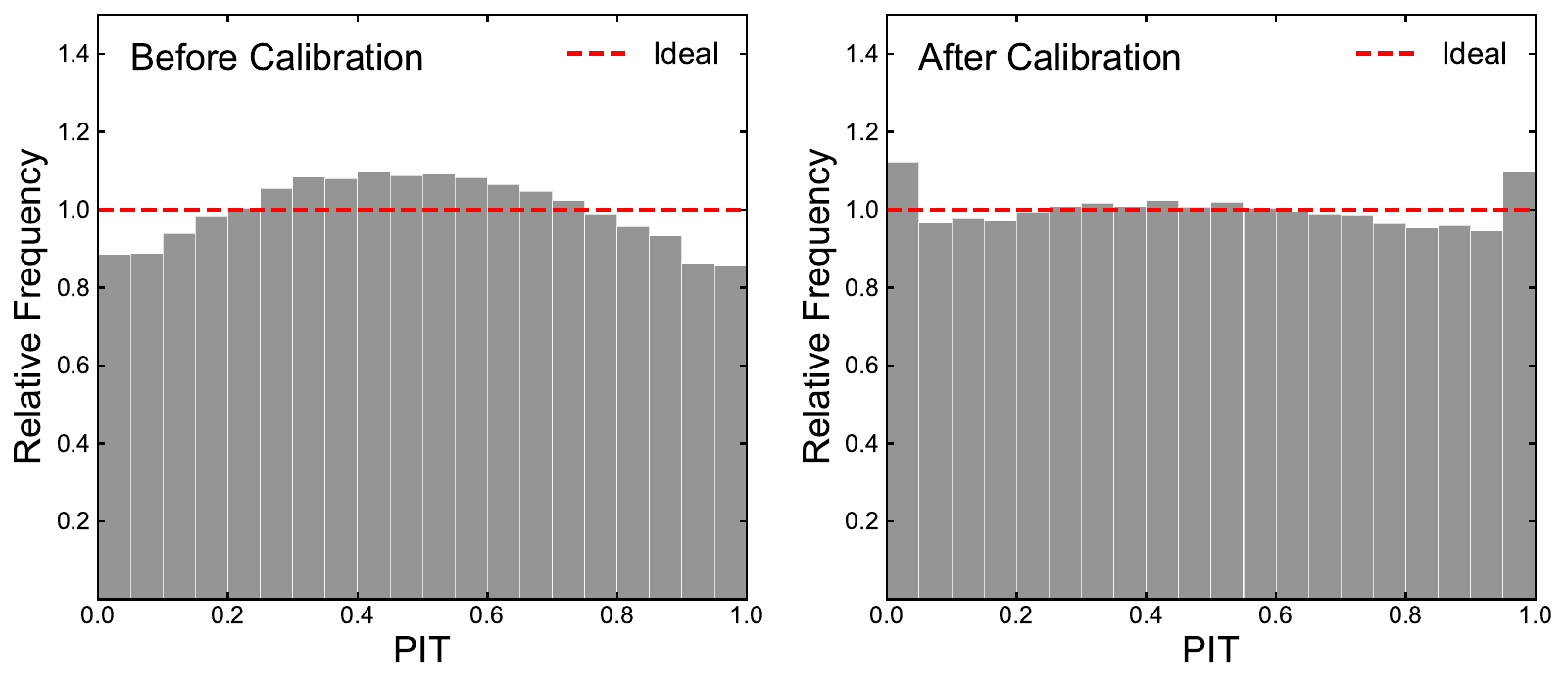}
\caption{PIT histograms for the LSDR10 test set before (left) and after (right) temperature scaling calibration.
The red dashed line indicates the ideal uniform distribution expected for perfectly calibrated PDFs. After calibration, the PIT distribution closely matches the uniform expectation.}
\label{fig:pit_calibration}
\end{figure}

We note that several alternative approaches for recalibrating photo-$z$ PDFs have been developed in the literature. \citet{bordoloi2010} introduced a recalibration method based on re-weighting the posterior using a reference spectroscopic sample. \citet{zhao2021} developed diagnostics for conditional density models that help identify local calibration failures across feature space. \citet{dey2021, dey2025} proposed feature-space and instance-wise recalibration techniques based on conditional density modeling. Our adoption of temperature scaling offers the advantage of simplicity—requiring optimization of only a single scalar parameter—while achieving near-uniform PIT distributions for our application (Figure~\ref{fig:pit_calibration}). For applications requiring stricter conditional calibration guarantees, the more flexible approaches cited above may provide additional benefits.

Figure~\ref{fig:pit_calibration} compares the PIT histograms before and after applying temperature scaling calibration.
The raw model outputs exhibit a slight inverted U-shaped distribution, indicating that the predicted PDFs are somewhat broader than necessary.
After temperature scaling calibration, the PIT distribution becomes nearly uniform, demonstrating that our calibrated PDFs provide reliable uncertainty estimates.

\begin{figure}[ht!]
\centering
\includegraphics[width=\textwidth]{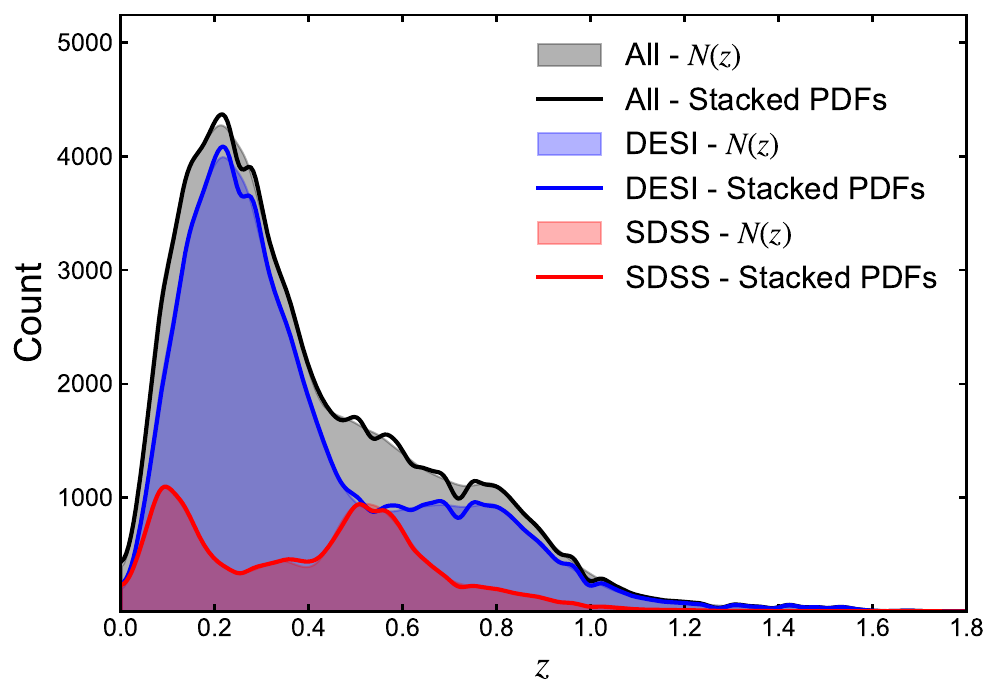}
\caption{Comparison of the true redshift distribution $N(z)$ (shaded regions) and the distribution from stacked PDFs (solid lines) for the LSDR10 test set, shown for all sources (black), the DESI subsample (blue), and the SDSS subsample (red).}
\label{fig:stacked_pdf}
\end{figure}

Finally, we examine the ensemble behavior of our PDFs by comparing the stacked photo-z distribution with the true redshift distribution.
Figure~\ref{fig:stacked_pdf} shows the true $N(z)$ (shaded regions) and the distribution from stacked PDFs (solid lines) for the full test set and subsamples from DESI and SDSS.
The bimodal structure of the SDSS subsample in Figure~\ref{fig:stacked_pdf} primarily reflects the combination of the SDSS Main Galaxy Sample at $z \lesssim 0.3$ and the BOSS LOWZ/CMASS samples at $z \sim 0.3$--$0.7$.
The overall agreement between the true and stacked-PDF distributions across all redshift ranges is broadly consistent with our PDFs capturing the main statistical properties of the sample.
We note that stacking individual PDFs is not an unbiased estimator of the true $N(z)$ \citep{leistedt2016, malz2021}; the comparison here serves as a visual consistency check rather than a rigorous cosmological $N(z)$ reconstruction.

\subsection{Comparison with Other Methods} \label{sec:comparison_ml}

To assess the performance of our NNC method relative to other commonly used machine learning approaches, we train and evaluate Random Forest (RF) \citep{breiman2001}, XGBoost \citep{chen2016}, and standard ANN regression models on LSDR10 using identical training, validation, and test splits.
For RF and XGBoost, we use the scikit-learn \citep{pedregosa2011} and XGBoost libraries with hyperparameters optimized via grid search on the validation set.
The ANN regression model shares the same architecture as our NNC model but outputs a single continuous redshift value trained with Huber loss.
As summarized in Table~\ref{tab:table1}, our NNC method achieves the best performance with $\sigma_{\mathrm{NMAD}} = 0.0153$ and $\eta = 0.50\%$, outperforming RF ($\sigma_{\mathrm{NMAD}} = 0.0278$), XGBoost ($\sigma_{\mathrm{NMAD}} = 0.0220$), and standard ANN regression ($\sigma_{\mathrm{NMAD}} = 0.0170$).
The NNC method achieves reductions of 45\%, 31\%, and 10\% in $\sigma_{\mathrm{NMAD}}$ over RF, XGBoost, and ANN regression, respectively.

We further compare our results with two benchmarks: the official LSDR10 photometric redshifts from \citet{zhou2023} and the recent catalog from \citet{wen2024a}.

First, compared to the official LSDR10 catalog by \citet{zhou2023}, who employed a Random Forest trained on earlier spectroscopic samples, our method demonstrates consistent improvements.
Using the aforementioned LSDR10 test set, our method achieves $\sigma_{\mathrm{NMAD}} = 0.0153$ and $\eta = 0.50\%$, compared to $\sigma_{\mathrm{NMAD}} = 0.0159$ and $\eta = 0.77\%$ for \citet{zhou2023} using $g,r,i,z_{\mathrm{m}},W1,W2$ bands, and $\sigma_{\mathrm{NMAD}} = 0.0186$ and $\eta = 0.67\%$ using $g,r,z_{\mathrm{m}},W1,W2$ bands.
Our results show a slight improvement over their $g,r,i,z_{\mathrm{m}},W1,W2$ configuration and a substantial improvement over their $g,r,z_{\mathrm{m}},W1,W2$ configuration, even though their models additionally incorporate morphological information as input features.
This advantage becomes pronounced for faint sources ($z_{\mathrm{m}} > 21$), where our method yields $\sigma_{\mathrm{NMAD}} = 0.0449$ and $\eta = 7.29\%$, versus $\sigma_{\mathrm{NMAD}} = 0.0512$ and $\eta = 8.97\%$ for \citet{zhou2023} ($g,r,i,z_{\mathrm{m}},W1,W2$).

Second, we compare our performance with that of \citet{wen2024a} (who used the nearest-neighbor algorithm) on the overlapping subset of the test set.
Our method exhibits superior accuracy and lower bias across all magnitude bins.
For the full sample, we reduce the scatter to $\sigma_{\mathrm{NMAD}} = 0.0150$ with an outlier fraction of $\eta = 0.38\%$, significantly outperforming \citet{wen2024a} who reported $\sigma_{\mathrm{NMAD}} = 0.0170$ and $\eta = 0.68\%$.
This performance advantage persists for faint sources ($z_{\mathrm{m}} > 21$), where we achieve $\sigma_{\mathrm{NMAD}} = 0.0372$ and $\eta = 4.61\%$, compared to $\sigma_{\mathrm{NMAD}} = 0.0396$ and $\eta = 5.79\%$ in \citet{wen2024a}.

For PS1DR2, previous photo-$z$ studies using optical bands alone achieved $\sigma_{\mathrm{NMAD}} \sim 0.026$ \citep{lee2021}.
In this work, we improve the PS1DR2 photo-$z$ precision to $\sigma_{\mathrm{NMAD}} = 0.0222$ by combining optical photometry with unWISE mid-infrared data.

In summary, the overall improvement over these existing catalogs and previous works is attributed to two key factors: the expanded, high-completeness training set provided by DESI DR1, and the robust uncertainty quantification and optimization capability of our CRPS-based NNC framework.

\section{Discussion} \label{sec:discussion}

\subsection{Advantages of the NNC Method} \label{sec:discuss_method}

As demonstrated in Section~\ref{sec:comparison_ml}, our NNC method achieves superior performance compared to commonly used machine learning approaches including Random Forest, XGBoost, and standard ANN regression.
On LSDR10, NNC achieves $\sigma_{\mathrm{NMAD}} = 0.0153$, representing reductions of 45\%, 31\%, and 10\% in $\sigma_{\mathrm{NMAD}}$ over RF, XGBoost, and ANN regression, respectively.
This performance advantage arises from two key aspects of our method: the CRPS loss function and the probabilistic output framework.

First, the CRPS loss function is naturally suited for photo-$z$ estimation.
CRPS operates on cumulative distribution functions and measures the integrated squared difference between the predicted CDF and a step function at the true redshift, a natural representation given the high precision of spectroscopic measurements.
This formulation imposes distance-sensitive penalties: errors in distant bins are penalized more heavily than errors in nearby bins, naturally respecting the ordinal structure of redshift.
Consequently, CRPS optimization guides the model to produce sharply peaked PDFs centered on the correct redshift, leading to lower $\sigma_{\mathrm{NMAD}}$ compared to standard ANN regression using Huber loss.

Second, the probabilistic output framework offers multiple advantages over single-value regression:
(1) It enables robust point estimates through expectation values, which naturally suppress the influence of low-probability tails and reduce sensitivity to outliers.
(2) It captures multi-modal posterior distributions arising from color-redshift degeneracies. As illustrated in Figure~\ref{fig:pdf_examples}, galaxies with ambiguous photometric colors may have multiple plausible redshift solutions—a scenario that traditional regression methods cannot represent.
(3) It provides well-calibrated uncertainty quantification. Combined with temperature scaling calibration, the predicted PDFs yield uniform PIT distributions (Figure~\ref{fig:pit_calibration}), indicating statistically reliable uncertainties essential for downstream cosmological analyses.

\subsection{DESI DR1: Contributions and Limitations} \label{sec:discuss_desi}

As characterized in Section~\ref{sec:spectro_properties}, DESI DR1 expands the spectroscopic training sample by a factor of $\sim$3 compared to SDSS alone, with critical improvements in both redshift range and magnitude coverage (Figure~\ref{fig:color_redshift_comparison}).
These improvements translate directly into better photo-$z$ performance: as shown in Figure~\ref{fig:nmad_vs_z}, DESI-trained models maintain $\sigma_{\mathrm{NMAD}} \lesssim 0.05$ out to $z \sim 1.6$, whereas SDSS-only models degrade rapidly beyond $z \sim 1$.

Despite these advances, several limitations warrant consideration.
On the one hand, DESI target selection prioritizes specific science goals over magnitude-limited sampling, causing LRGs, ELGs, and BGS to occupy distinct regions of color-magnitude space while leaving intermediate-color populations underrepresented.
For sources in these sparsely sampled regions, the model must extrapolate rather than interpolate, potentially degrading accuracy.
On the other hand, DESI DR1 covers $\sim$9,000 deg$^2$, leaving portions of the sky with limited spectroscopic coverage, but this limitation will be alleviated as the complete DESI survey expands to $\sim$40 million spectra over 14,000 deg$^2$.

\subsection{Why LSDR10 Outperforms PS1DR2} \label{sec:discuss_photometry}

The most obvious difference between LSDR10 and PS1DR2 is wavelength coverage.
LSDR10 incorporates WISE W1 and W2 mid-infrared photometry, which traces stellar mass and provides useful indirect constraints that help mitigate color-redshift degeneracies, particularly for red galaxies at $z > 0.5$ where the 4000~\AA\ break shifts beyond the optical window.
PS1DR2, with only optical $g,r,i,z_{\mathrm{m}},y$ bands, cannot distinguish between low-redshift red galaxies and high-redshift blue galaxies that share similar optical colors.

To address this limitation, we augmented PS1DR2 with unWISE infrared photometry.
As shown in Table~\ref{tab:table1}, adding W1/W2 bands significantly improves PS1DR2 performance from $\sigma_{\mathrm{NMAD}} = 0.0283$ to $0.0222$, representing a $\sim$22\% reduction in scatter.
We note that a significant wavelength gap ($\sim 1$--$3.4\,\mu$m) exists between the reddest optical bands (e.g., $z_{\mathrm{m}}$ or $y$) and WISE W1, so the 4000~\AA\ break cannot be directly tracked as it redshifts into the near-infrared. Rather than tracing this spectral feature, WISE provides indirect redshift constraints through the long optical-to-infrared color baseline (e.g., $z_{\mathrm{m}} - W1$) and the rest-frame near-infrared luminosity, which correlates with stellar mass and galaxy type. Filling this wavelength gap with near-infrared photometry, for example from future Euclid $YJH$ bands, would further improve photo-$z$ precision \citep{ilbert2008, desprez2020}.
However, PS1DR2 + unWISE still exhibits $\sigma_{\mathrm{NMAD}}$ approximately 1.45 times larger than LSDR10 ($\sigma_{\mathrm{NMAD}} = 0.0153$), even though both surveys now have infrared coverage.
This residual performance gap suggests that photometric depth plays a crucial role beyond wavelength coverage alone.
LSDR10 achieves $5\sigma$ point-source depths of approximately 24.7, 23.9, and 23.0 mag in $g$, $r$, and $z_{\mathrm{m}}$ bands, respectively, compared to 23.3, 23.2, and 22.3 mag for PS1DR2.
The deeper photometry provides more precise color measurements and reduces photometric uncertainties.

\begin{figure}[ht!]
\centering
\includegraphics[width=\textwidth]{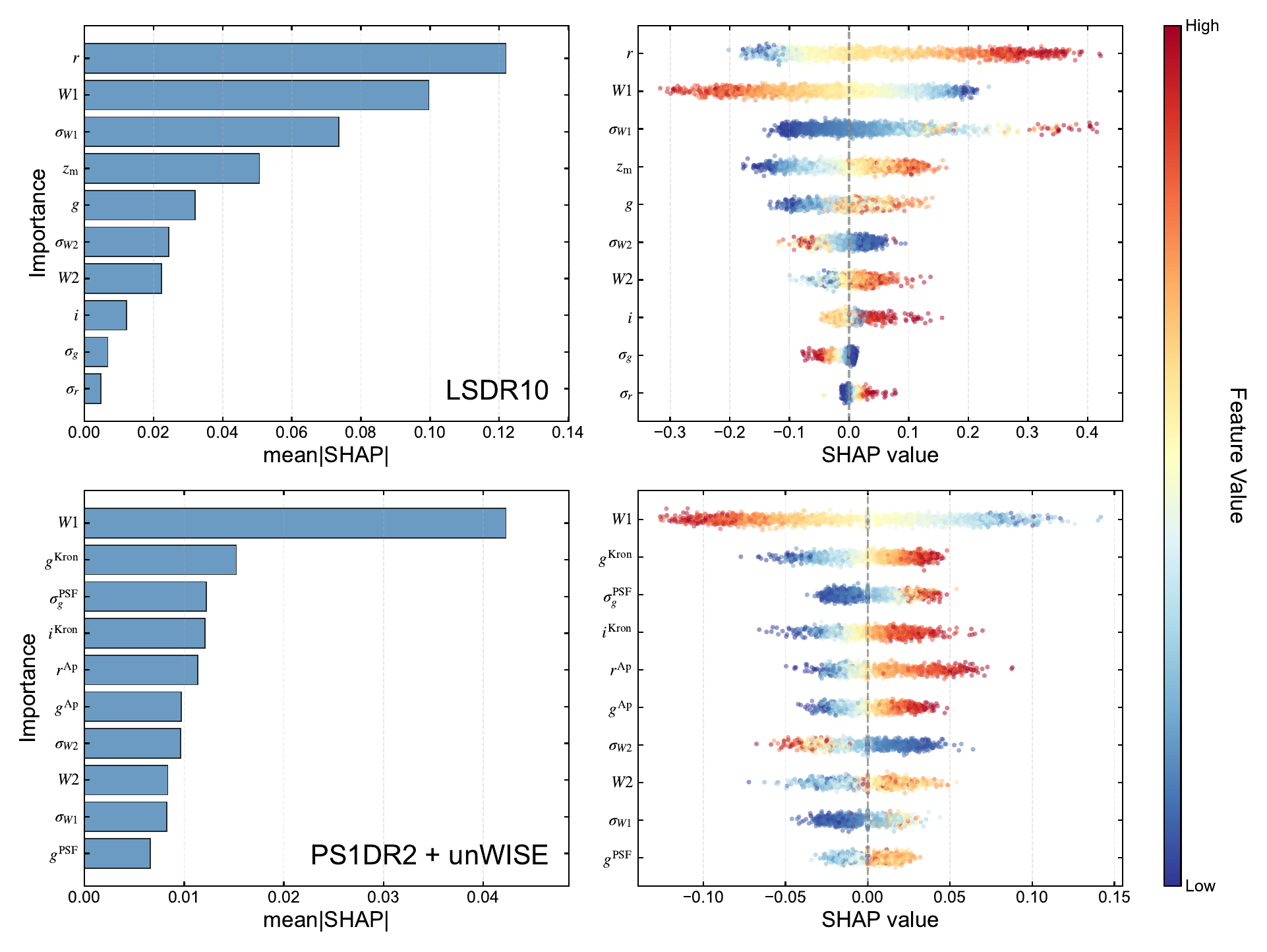}
\caption{SHAP analysis for the LSDR10 (upper) and PS1DR2 + unWISE (lower) models. Left panels show the mean absolute SHAP value, mean($|\mathrm{SHAP}|$), for each feature as a quantitative measure of its overall importance. Right panels display per-galaxy SHAP values for the top-ranked features, where each point represents one test galaxy. The horizontal position indicates the feature's contribution to the predicted redshift (positive values increase and negative values decrease the prediction relative to the baseline). The color encodes the measured value of the feature itself: for photometric magnitudes, red corresponds to faint (high-magnitude) sources and blue to bright (low-magnitude) sources; for magnitude errors, red indicates large uncertainties.}
\label{fig:shap_feature_importance_comparison}
\end{figure}

To further investigate the impact of photometric depth on model photo-$z$ prediction, we employ the SHAP (SHapley Additive exPlanations) method \citep{lundberg2017} to quantitatively assess how different photometric features contribute to redshift estimation.
SHAP provides a principled approach to fairly attribute a model's prediction to its input features by computing the average marginal contribution of each feature across all possible feature combinations.
We quantify the overall importance of each feature using mean($|\mathrm{SHAP}|$), the mean absolute SHAP value across all test galaxies (Figure~\ref{fig:shap_feature_importance_comparison}, left panels). The right panels show the per-galaxy SHAP values: positive (negative) values indicate that a feature pushes the predicted redshift higher (lower), while the color encodes the feature's own measured value. For photometric magnitude features, red points correspond to faint sources and blue points to bright sources; for magnitude error features, red indicates larger measurement uncertainties.

As shown in Figure~\ref{fig:shap_feature_importance_comparison}, the LSDR10 and PS1DR2 + unWISE models exhibit markedly different feature utilization patterns.
For LSDR10, the feature importance is led by the optical $r$ band, followed by the infrared $W1$ band and its associated error $\sigma_{W1}$, then by the optical $z_{\mathrm{m}}$ and $g$ bands.
The SHAP values span a wide range from approximately $-0.3$ to $+0.4$, indicating that individual photometric bands provide strong constraints on redshift through SED features.

In contrast, the PS1DR2 + unWISE model shows a notably different pattern.
The infrared $W1$ band emerges as the most important feature, followed by $g^{\mathrm{Kron}}$ and the magnitude error $\sigma_g^{\mathrm{PSF}}$, then by other optical magnitudes from different aperture systems ($i^{\mathrm{Kron}}$, $r^{\mathrm{Ap}}$, $g^{\mathrm{Ap}}$).
Notably, magnitude errors ($\sigma_g^{\mathrm{PSF}}$, $\sigma_{W2}$) rank prominently among the top features, and the SHAP value range is compressed to approximately $-0.10$ to $+0.15$, roughly three times narrower than LSDR10.
This shift in weight from photometric magnitudes to measurement uncertainties and multi-aperture differences indicates that the model relies increasingly on indirect proxies when single-band signal-to-noise ratios are insufficient.
This explains why deeper photometry, not just broader wavelength coverage, is essential for achieving high photo-$z$ precision.

\section{Photometric Redshift Catalog} \label{sec:catalog}

We construct a unified photometric redshift catalog that combines LSDR10 and PS1DR2 to maximize sky coverage while optimizing photo-$z$ precision for each source.
The DESI Legacy Imaging Surveys and Pan-STARRS have complementary sky coverage and photometric properties: LSDR10 provides deeper photometry with integrated WISE infrared bands but is limited to approximately 20,000 deg$^2$, while PS1DR2 covers a larger area of approximately 30,000 deg$^2$ (dec $> -30^{\circ}$) but with shallower optical depth.
Additionally, native LSDR10 $i$-band photometry is available only for a subset of sources, primarily within the DECaLS footprint, so we adopt different models depending on the available band coverage.

To leverage the strengths of each survey and provide optimal photo-$z$ estimates for all sources, we adopt a hierarchical strategy with five priority levels based on available photometry (performance metrics are summarized in Table~\ref{tab:table1}):
\begin{enumerate}
    \item \textbf{Priority 1: LSDR10 full bands ($g,r,i,z_{\mathrm{m}},W1,W2$).}
    Sources with valid LSDR10 $g,r,i,z_{\mathrm{m}},W1,W2$ photometry.

    \item \textbf{Priority 2: LSDR10 + PS1DR2 $i$-band ($g,r,z_{\mathrm{m}},W1,W2$ + PS1DR2 $i$).}
    Sources with valid LSDR10 $g,r,z_{\mathrm{m}},W1,W2$ photometry but without usable native LSDR10 $i$-band measurements, and with successful PS1DR2 $i$-band cross-matches.

    \item \textbf{Priority 3: LSDR10 without $i$-band ($g,r,z_{\mathrm{m}},W1,W2$).}
    Sources with valid LSDR10 $g,r,z_{\mathrm{m}},W1,W2$ photometry but without usable native LSDR10 $i$-band measurements or PS1DR2 $i$-band cross-matches.

    \item \textbf{Priority 4: PS1DR2 + unWISE ($g,r,i,z_{\mathrm{m}},y$ + $W1W2$).}
    Sources outside the LSDR10 footprint with valid PS1DR2 optical photometry and successful unWISE cross-matches.

    \item \textbf{Priority 5: PS1DR2 optical only ($g,r,i,z_{\mathrm{m}},y$).}
    Sources with PS1DR2 optical photometry but without valid unWISE matches.
\end{enumerate}
For sources in the overlapping regions between surveys, duplicates are identified through cross-matching within a $1^{\prime\prime}$ radius, and the higher-priority model is applied.

The final catalog contains about 552 million galaxies, with 341, 34, 149, 13, and 15 million sources in Priority 1--5, respectively.
The catalog is publicly available in FITS format\footnote{\url{https://doi.org/10.5281/zenodo.18410730}. See this record for the current version (v2.0), which uses DESI Legacy Imaging Surveys DR11 and replaces the DR10 priority levels with \texttt{survey} and \texttt{model} columns.}, where detailed column descriptions are provided. The source code used in this work is publicly available on GitHub\footnote{\url{https://github.com/tdccccc/photoz-nnc}}. The catalog includes source coordinates (RA, Dec), photometric redshift estimates (expectation value, mode, and uncertainty derived from the PDF), priority levels, and survey flags indicating whether the source is detected in LSDR10 and/or PS1DR2.
To balance completeness with practical usability, the data release includes a 40-bin redshift PDF obtained by resampling the native 400-bin model output. Releasing all 400 bin probabilities for every source would substantially increase the file size, whereas the 40-bin representation is more convenient for download and routine use while still preserving the overall probability distribution.

\section{Summary} \label{sec:summary}

We develop an NNC method for photometric redshift estimation and apply it to LSDR10 and PS1DR2.
The model is trained on an unprecedented spectroscopic sample from DESI DR1 and SDSS DR19, which provides relatively broad coverage across redshift and color-magnitude space.
By discretizing the redshift space into bins and optimizing the CRPS, our method produces well-calibrated PDFs that naturally capture multi-modal posteriors and provide reliable uncertainty quantification.
Our method achieves high-precision photometric redshifts with $\sigma_{\mathrm{NMAD}} = 0.0153$ and $\eta = 0.50\%$ on LSDR10, and $\sigma_{\mathrm{NMAD}} = 0.0222$ and $\eta = 0.34\%$ on PS1DR2 when combined with unWISE infrared photometry.

Our main conclusions are as follows:
\begin{enumerate}
    \item The NNC method outperforms Random Forest, XGBoost, and standard ANN regression on both surveys. The CRPS loss function improves redshift accuracy and produces well-calibrated PDFs suitable for downstream cosmological analyses.

    \item DESI DR1 dramatically expands spectroscopic coverage to $z > 1$ through LRG and ELG targets, while the combination with SDSS provides more complete sampling in color-magnitude space, yielding optimal photo-$z$ performance across the full redshift range.

    \item Adding unWISE W1/W2 infrared photometry to PS1DR2 significantly reduces scatter and outlier fraction,
    demonstrating that mid-infrared coverage provides useful indirect constraints that help improve photo-$z$ precision. The remaining performance gap with LSDR10 is attributed to the shallower optical photometric depth of PS1DR2.


    \item We construct a unified photometric redshift catalog covering $>$30,000 deg$^2$ by combining LSDR10 and PS1DR2 with a hierarchical five-level strategy that assigns each source to the optimal model based on available photometry, maximizing both sky coverage and photo-$z$ precision.
\end{enumerate}

Our NNC framework is survey-agnostic and can be directly applied to next-generation imaging surveys such as CSST, Euclid, and LSST by redefining the input features to match the available photometric bands (e.g., CSST's $NUV, u, g, r, i, z_{\mathrm{m}}, y$ bands; Euclid's $Y, J, H$ combined with ground-based optical photometry from LSST). However, the primary challenge for these surveys lies in the spectroscopic training data. Our current DESI+SDSS sample provides relatively broad coverage in redshift and color space up to $z \lesssim 1.6$, whereas future weak-lensing-oriented imaging surveys will routinely detect galaxies at $z > 2$ with magnitudes well beyond current spectroscopic limits. Building representative training sets at these redshifts will likely require future spectroscopic programs such as DESI-II and 4MOST \citep{jong2019}. In regimes where spectroscopy remains sparse, hybrid approaches combining data-driven methods with template-based priors or transfer learning may be necessary to ensure robust photo-$z$ estimation for weak lensing and other precision cosmology applications.

\begin{acknowledgments}
We thank Liang Jing for helpful discussions and suggestions on the Pan-STARRS data processing.

J.-Q.X. is supported by the National Natural Science Foundation of China, under grant No. 12473004. Z.-L.W. is supported by the National Natural Science Foundation of China, under grant No. 12073036. This work is also supported by the China Manned Space Program, grant Nos. CMS-CSST-2025-A01 and CMS-CSST-2025-A04; and the Fundamental Research Funds for the Central Universities.

Funding for the Sloan Digital Sky Survey V has been provided by the Alfred P. Sloan Foundation, the Heising-Simons Foundation, the National Science Foundation, and the Participating Institutions. SDSS acknowledges support and resources from the Center for High-Performance Computing at the University of Utah. SDSS telescopes are located at Apache Point Observatory, funded by the Astrophysical Research Consortium and operated by New Mexico State University, and at Las Campanas Observatory, operated by the Carnegie Institution for Science. The SDSS web site is \url{www.sdss.org}.

SDSS is managed by the Astrophysical Research Consortium for the Participating Institutions of the SDSS Collaboration, including the Carnegie Institution for Science, Chilean National Time Allocation Committee (CNTAC) ratified researchers, Caltech, the Gotham Participation Group, Harvard University, Heidelberg University, The Flatiron Institute, The Johns Hopkins University, L'Ecole polytechnique f\'{e}d\'{e}rale de Lausanne (EPFL), Leibniz-Institut f\"{u}r Astrophysik Potsdam (AIP), Max-Planck-Institut f\"{u}r Astronomie (MPIA Heidelberg), Max-Planck-Institut f\"{u}r Extraterrestrische Physik (MPE), Nanjing University, National Astronomical Observatories of China (NAOC), New Mexico State University, The Ohio State University, Pennsylvania State University, Smithsonian Astrophysical Observatory, Space Telescope Science Institute (STScI), the Stellar Astrophysics Participation Group, Universidad Nacional Aut\'{o}noma de M\'{e}xico, University of Arizona, University of Colorado Boulder, University of Illinois at Urbana-Champaign, University of Toronto, University of Utah, University of Virginia, Yale University, and Yunnan University.

This research used data obtained with the Dark Energy Spectroscopic Instrument (DESI). DESI construction and operations is managed by the Lawrence Berkeley National Laboratory. This material is based upon work supported by the U.S. Department of Energy, Office of Science, Office of High-Energy Physics, under Contract No. DE--AC02--05CH11231, and by the National Energy Research Scientific Computing Center, a DOE Office of Science User Facility under the same contract. Additional support for DESI was provided by the U.S. National Science Foundation (NSF), Division of Astronomical Sciences under Contract No. AST-0950945 to the NSF's National Optical-Infrared Astronomy Research Laboratory; the Science and Technology Facilities Council of the United Kingdom; the Gordon and Betty Moore Foundation; the Heising-Simons Foundation; the French Alternative Energies and Atomic Energy Commission (CEA); the National Council of Humanities, Science and Technology of Mexico (CONAHCYT); the Ministry of Science and Innovation of Spain (MICINN), and by the DESI Member Institutions: www.desi.lbl.gov/collaborating-institutions. The DESI collaboration is honored to be permitted to conduct scientific research on I'oligam Du'ag (Kitt Peak), a mountain with particular significance to the Tohono O'odham Nation. Any opinions, findings, and conclusions or recommendations expressed in this material are those of the author(s) and do not necessarily reflect the views of the U.S. National Science Foundation, the U.S. Department of Energy, or any of the listed funding agencies.

The Pan-STARRS1 Surveys (PS1) have been made possible through contributions of the Institute for Astronomy, the University of Hawaii, the Pan-STARRS Project Office, the Max-Planck Society and its participating institutes, the Max Planck Institute for Astronomy, Heidelberg and the Max Planck Institute for Extraterrestrial Physics, Garching, The Johns Hopkins University, Durham University, the University of Edinburgh, Queen's University Belfast, the Harvard-Smithsonian Center for Astrophysics, the Las Cumbres Observatory Global Telescope Network Incorporated, the National Central University of Taiwan, the Space Telescope Science Institute, the National Aeronautics and Space Administration under Grant No. NNX08AR22G issued through the Planetary Science Division of the NASA Science Mission Directorate, the National Science Foundation under Grant No. AST-1238877, the University of Maryland, and Eotvos Lorand University (ELTE), the Los Alamos National Laboratory, and the Gordon and Betty Moore Foundation.

This publication makes use of data products from the Wide-field
Infrared Survey Explorer, which is a joint project of the University of California, Los Angeles, and the Jet Propulsion Laboratory/California Institute of Technology, funded by the National Aeronautics and Space Administration.

\end{acknowledgments}

\bibliography{ref}{}
\bibliographystyle{aasjournalv7}

\end{document}